\documentclass[aip, amsmath,amssymb,reprint]{revtex4-2}

\usepackage{graphicx}
\usepackage{dcolumn}
\usepackage{bm}
\usepackage{siunitx}
\usepackage{hyperref}
\hypersetup{colorlinks = true,linkcolor=blue,
     filecolor=blue,
     urlcolor=blue,
     citecolor = blue, pdfauthor=author}

\begin{document}

\preprint{AIP/123-QED}

\title{Longitudinal-Twist Wave Converter based on Chiral Metamaterials}
\author{Brahim Lemkalli}
\email{brahim.lemkalli@edu.umi.ac.ma}
\affiliation{Laboratory for the Study of Advanced Materials and Applications, Department of Physics, Moulay Ismail University, B.P. 11201, Zitoune, Meknes, Morocco.}

\author{Muamer Kadic}
\affiliation{Institut FEMTO-ST, UMR 6174, CNRS, Universit\'{e} de Bourgogne Franche-Comt\'{e}, 25000 Besan\c{c}on, France}

\author{Youssef El Badri}
\affiliation{Laboratory of optics, information processing, Mechanics, Energetics and Electronics, Department of Physics, Moulay Ismail University, B.P. 11201, Zitoune, Meknes, Morocco}

\author{S\'{e}bastien Guenneau}
\affiliation{UMI 2004 Abraham de Moivre-CNRS, Imperial College London, SW7 2AZ, UK}

\author{Abdellah Mir}
\affiliation{Laboratory for the Study of Advanced Materials and Applications, Department of Physics, Moulay Ismail University, B.P. 11201, Zitoune, Meknes, Morocco.}

\author{Younes Achaoui}
\affiliation{Laboratory for the Study of Advanced Materials and Applications, Department of Physics, Moulay Ismail University, B.P. 11201, Zitoune, Meknes, Morocco.}

\date{\today}

\begin{abstract}
Advances in material architectures have enabled endowing materials with exotic attributes not commonly available in the conventional realm of mechanical engineering. Twisting, a mechanism whereby metamaterials are used to transform static axial load into twist motion, is of particular interest to this study. Herein, computations based on the finite element method, corroborated by a theoretical approach derived from applying Lagrange's equations to a monoatomic spring-mass system, are employed to explore the longitudinal-twist (L-T) conversion exhibited by a chiral tetragonal-beam metamaterial. Firstly, we perform an eigenvalue analysis taking into account the polarization states to highlight the potential contribution of the longitudinal mode in the L-T conversion. We contrast the twisting behavior of the chiral cell with that of other homogeneous medium, octagonal-tube, and non-chiral cells. Moreover, we demonstrate the influence of the cell's chirality on the L-T conversion using both time-domain and frequency-domain studies. The findings indicate that at least a portion of the longitudinally propagating wave is transformed into twist throughout a broad frequency range and even quasi-totally converted at distinct frequencies.
\end{abstract}

\maketitle

\section{Introduction}
\begin{figure}
    \centering
    \includegraphics[width=7.5cm,angle=0]{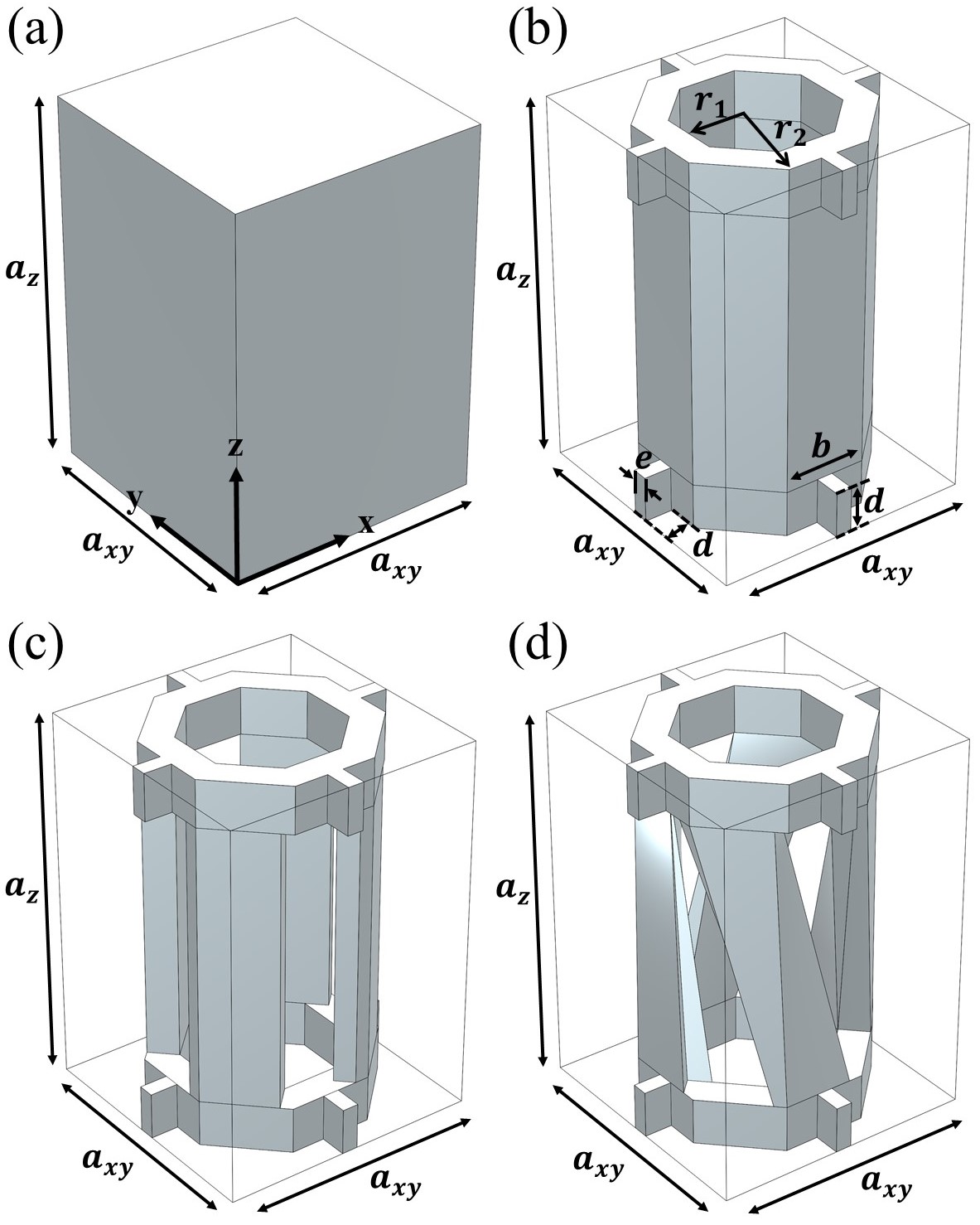}
    \caption{\label{Figure 1}Schematics of the tetragonal-beam unit cells. (a) The homogeneous medium. (b) The octagonal tube comprised of the inscribed $r_{1}=2.31$ \si{mm} and circumscribed $r_{2}=3.56$ \si{mm} circles, and $8$ arms of length equal height $d=1.5$ \si{mm} and width $e=0.5$ \si{mm} attached to two octagonal plates in the $xy$-plane. (c) The non-chiral made up of vertical bars fixed to two octagonal plates. (d) The chiral unit cell composed of four inclined bars fixed to two octagonal plates. The tetragonal-beam with lattice parameters $a_{z}=12$ \si{mm}, $a_{xy}=8.4$ \si{mm}, and $b=2.2$ \si{mm}.}
\end{figure}

\indent Modern manufacturing techniques have facilitated the development of novel metamaterials with remarkable properties not previously observed in conventional materials \cite{kadic20193d, wu2019mechanical}. These composite materials have benefited a wide range of engineering fields, including optics \cite{soukoulis2010optical, mccall2018roadmap}, acoustics \cite{cummer2016controlling, ma2016acoustic, fleury2016floquet}, thermodynamics \cite{peralta2020brief, li2021transforming}, mechanics \cite{yu2018mechanical, surjadi2019mechanical} and even multiphysics \cite{kadic2015experiments,kadic2015hall,qu2017experiments}. The rapid development of mechanical metamaterials has resulted in the emergence of numerous exotic phenomena, some of which have static attributes: negative Poisson's ratio \cite{ren2018auxetic}, programmable behavior \cite{florijn2014programmable}, and compression-twist conversion\cite{frenzel2017three}; others can achieve time-varying properties, i.e., dynamic ones. Consequently, instruments capable of manipulating the propagation of elastic \cite{buckmann2014elasto, zhu2014negative}, seismic \cite{brule2014experiments}, and acoustic waves \cite{wang2011large} are swiftly evolving. This will lead to the ultimate goal of attenuating, filtering, harnessing, and guiding mechanical waves.\newline
\indent Chiral twist metamaterials, which allow mechanical axial load to be transformed into twist motion in a static configuration, were first introduced in 2017 \cite{frenzel2017three}. Since then, major optimizations have been carried out in order to improve their twist behavior \cite{zhong2019novel, zheng2019novel, frenzel2021large, lemkalli2022mapping}. In particular, this category of metamaterials was proved to be capable of converting linearly polarized waves into elliptical transverse waves, a process generally referred to as acoustic activity \cite{frenzel2019ultrasound, reinbold2019rise, chen2020mapping}, which is analogous to optical activity \cite{plum2009metamaterials}. It is noteworthy that the occurrence of this phenomenon is manifested in the degeneration lifting of the transverse modes, which are directly related to the unit cell's chirality. The question, however, arises whether there is any relevance to the combination of longitudinal and twist waves that are inextricably linked to twist metamaterial behavior?\newline
\indent Researchers have been looking at the feasibility of transforming longitudinal waves into twist waves using various mechanisms, such as rotary ultrasonic motors \cite{tsujino1996load}, ultrasonic step horns \cite{harkness2012coupling}, sources that can regulate the longitudinal-twisting vibration \cite{asami2016longitudinal}, and the combination of several types of elastic waves \cite{de2022tailored}. In this context, the elastic spiral phase pipe that is analogous to its optical counterpart is used to convert axis-symmetric longitudinal guided modes in pipes into arbitrary non-axis-symmetric flexural modes \cite{chaplain2022elastic}. Furthermore, engineered structures that exhibit L-T conversion have also been reported. For instance, the conversion of longitudinal polarization into a hybrid wave representing both longitudinal and twisting modes has been proved by the Origami Kresling converter \cite{xu2022origami}. Generally, the L-T conversion effect is utilized for a wide range of ultrasonic applications, such as welding and drilling systems \cite{wang2022design, li2022modeling}.\newline
\indent We present in this paper an L-T converter based on a twist metamaterial formed by chiral tetragonal unit cells. The unit cell consists of two octagonal plates with $r_1$ and $r_2$ being the radii of the inscribed and circumscribed circles, respectively. These two octagonal plates are separated by a distance of $a_z$ and connected vertically by four inclined bars. In order to highlight the L-T conversion effect in the proposed unit cell, we compare numerically the twist behavior of four structures: the homogeneous medium unit cell, a unit cell with an octagonal tube shape, a non-chiral unit cell, and a chiral unit cell (\autoref{Figure 1}(a), (b), (c), and (d)). Finally, we delineate a theoretical model that describes the twist behavior of the chiral unit cell and compare the obtained eigenvalues to the numerical results.\newline

\section{Numerical and Theoretical methods}
\subsection{Modeling details}
\begin{figure*}
    \centering
    \includegraphics[width=16.0cm,angle=0]{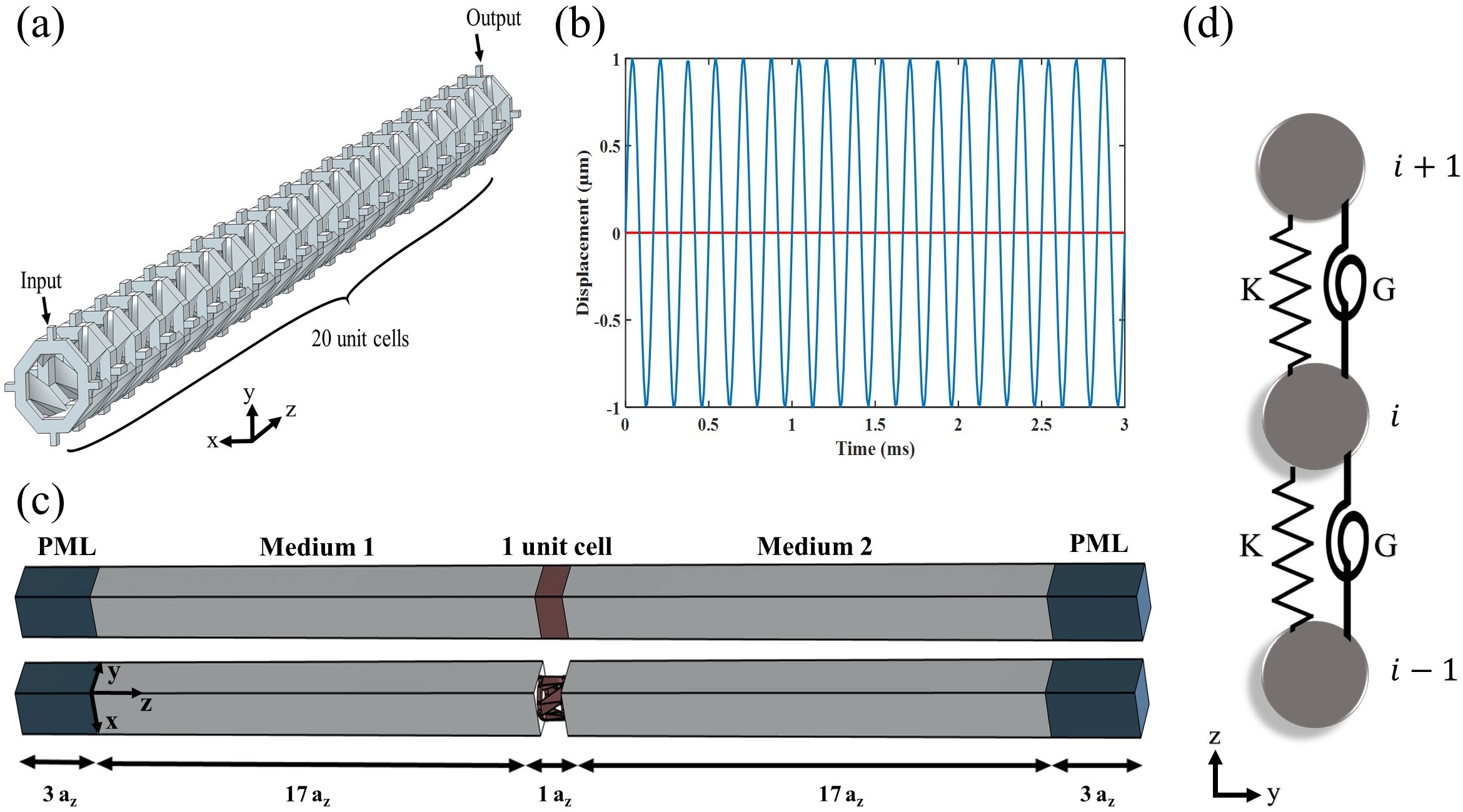}
    \caption{\label{Figure 2}(a) The array consists of 20 chiral unit cells. (b) The displacements triggered by the longitudinal excitation are showed as functions of time: the $u_x$ (red) and $u_z$ (blue). (c) The design of the geometries used in the harmonic study shows the unit cell and its adjacent media with PML at the borders. (d) Illustration of the spring-mass system. }
\end{figure*}
\indent We led a series of numerical simulations based on the finite element method using the commercial software COMSOL Multiphysics. During the study, we considered the constituent material to be Acrylonitrile Butadiene Styrene (ABS), which has the following mechanical properties: a Young's modulus $E=2.66$ \si{GPa}, a density $\rho=1020$ \si{kg/m^3} and a Poisson's ratio $\nu =0.4$.\newline
\indent The purpose is to investigate the effect of chirality on the conversion of longitudinal waves into twist waves and the coupling between the two waves. Accordingly, we accounted for the linear elastic nature of the material by implementing a numerical model based on the weak formalism of the time-harmonic Navier equation, via:
\begin{equation}\label{Eq01}
    \frac{E}{2(1+\nu)}(\frac{1}{(1-2\nu)}\nabla (\nabla.\textbf{u})+\nabla^2 \textbf{u})=-\omega^2\rho\textbf{u}
\end{equation}
where $\textbf{u}=(u_x, u_y, u_z)$ is the displacement vector and $\omega$ is the angular wave frequency.\newline
\indent First, we computed their phononic dispersion relations, assuming that the cells are repeated periodically along the $z$-direction and finite in the $xy$-plane. We applied the Floquet-Bloch boundary conditions along the $z$-direction as:
\begin{equation}\label{eq02}
    \textbf{u}(z+a_{z})=\textbf{u}(z)e^{ika_{z}},
\end{equation}
where ${\bf k}=(0, 0, k)$ is the reduced Bloch wavevector and $\textbf{u}(z)$ is the displacement vector.\newline
\indent We depicted the dispersion relation in the first irreducible Brillouin zone ($\Gamma Z$), where $\Gamma=(0, 0, 0)$ and $ Z=(0, 0, \pi/a_z)$. To delineate the L-T conversion effect, we computed the dispersion curves taking into account the polarization weighting using the following expression:
\begin{equation}\label{eq03}
    p_{z}=\frac{\iiint_V |u_{z}|\; dxdydz}{\iiint_V \sqrt{|u_{x}|^2+ |u_{y}|^2+ |u_{z}|^2}\; dxdydz},
\end{equation}
where $V$ is the total volume of each unit cell $|u_i|_{i=x, y, z}=\sqrt{u_i u_i^*}$ is the modulus of displacement vector, with $u_i^*$ the complex conjugate of $u_i$.\newline
\indent After examining the corresponding eigenmodes of the metamaterial, the L-T conversion is investigated through the simulation of time-varying wave propagation in an array of $20$ cells stacked along the $z$-direction for each of the four configurations, as portrayed in \autoref{Figure 2}(a). \autoref{Figure 2}(b) demonstrates that the structures were subjected to a sinusoidal longitudinal wave with a frequency of $6$ \si{kHz} and an amplitude of $1$ \si{\mu m} along the $z$-direction, with a free constraint on the $x$ and $y$ sides. This creates a purely longitudinal excitation wave. On the output of the $20$ cells, the $u_z$ and $u_x$ displacements are recorded as a function of time.\newline
\indent Since the L-T conversion effect in the chiral cell is dependent on the frequency of excitation, a harmonic study is also performed for all the configurations to identify their corresponding responses. As illustrated in \autoref{Figure 2}(c), medium $1$ is excited using a $1$ \si{\mu m} amplitude elastic wave along the $z$-direction with the $x$ and $y$ sides are in free constraint. Perfectly Matched Layers (PML) are used to gradually decrease the amplitude of the elastic wave before they reach the borders of medium $2$, thereby, eliminating reflections at the structure's boundaries. To quantify the conversion rate, we defined the longitudinal $R_L$ and twist $R_T$ conversion rates as follows:
\begin{equation}\label{eq04}
    R_{L}=\frac{\iiint_{V_2} |u_{z}|\; dxdydz}{\iiint_{V_2} \sqrt{|u_{x}|^2+ |u_{y}|^2+ |u_{z}|^2}\; dxdydz},
\end{equation}
\begin{equation}\label{eq05}
    R_{T}=\frac{\iiint_{V_2} \sqrt{|u_{x}|^2+ |u_{y}|^2}\; dxdydz}{\iiint_{V_2} \sqrt{|u_{x}|^2+ |u_{y}|^2+ |u_{z}|^2}\; dxdydz},
\end{equation} 
where $|u_i|$ is the modulus of the displacement vector, $V_2$ is the volume of medium $2$ and $R_L$, $R_T$ are the longitudinal and twist ratios, respectively.

\subsection{Theoretical model}
\indent In order to provide an theoretical model to describe the longitudinal-twist conversion, we reiterate that the chiral cell undergoes both axial deformation and twist mechanism. Accordingly, we proposed a monoatomic analytic model predicated upon the spring-mass systems, as illustrated in \autoref{Figure 2}(d). The mass (numbered $i$) is connected to the two neighboring masses ($i+1$ and $i-1$) separated by a distance $a_z$. The spring-mass equivalent model has two degrees of freedom: translations along the $z$-direction, which are indicated by the displacement $u_z$, and rotations around the $z$-axis, which are indicated by $\Phi$. In accordance with Lagrange's theory, the dynamic differential equations are as follows \cite{suiker2001comparison,pichard2014localized, bergamini2019tacticity, park2022chiral}:
\begin{equation}\label{eq06}
    M\frac{\partial^2 u_z}{\partial t^2}=K(u_z^{i+1}+u_z^{i-1}-2u_z^{i})\\-K\xi(\Phi^{i+1}+\Phi^{i-1}+2\Phi^{i}),
\end{equation}
\begin{multline}\label{eq07}
    J\frac{\partial^2 \Phi^i}{\partial t^2}=G(\Phi^{i+1}+\Phi^{i-1}-2\Phi^{i})\\+K\xi(u_z^{i+1}+u_z^{i-1}-2u_z^{i})\\-K\xi(\Phi^{i+1}-\Phi^{i-1}),
\end{multline}
where M is the mass, J is the moment of inertia, K is the stiffness, G is the rotational stiffness and $\xi$ is the longitudinal-twist coupling coefficient.
Floquet-Bloch periodicity conditions were applied to obtain the vibration parameters of the $i^{th}$ mass as follows: ${u^{i+1}=u^ie^{-jk_za_z}}$, ${u^{i-1}=u^{i}e^{jk_za_z}}$, ${\Phi^{i+1}=\Phi^ie^{-jk_za_z}}$, and ${\Phi^{i-1}=\Phi^{i}e^{jk_za_z}}$. By substituting these equations into Equations \ref{eq06} and \ref{eq07}, the differential equations for the mass $i$ can be obtained:
\begin{equation}\label{eq08}
    M\frac{\partial^2 u_z^i}{\partial t^2}=2K(\text{cos}(k_za_z)-1)u_z^i\\+2jK\xi \;\text{sin}(k_za_z)\Phi^i,
\end{equation}
\begin{multline}\label{eq09}
    J\frac{\partial^2 \Phi^i}{\partial t^2}=-2jK\xi \; \text{sin}(k_za_z) u_z^i\\ + 2G((\text{cos}(k_za_z)-1)\\-\xi^2(\text{cos}(k_za_z)+1))\Phi^i.
\end{multline}
\indent In order to determine the eigenvalues of the longitudinal and twist modes, we compute the roots of the characteristic polynomial $det(A(k_z)-B\omega^2\mathbb{I})$, where: 
\begin{multline}\label{eq10}
A(k_z)=\left (\begin{array}{cc} {2K(\text{cos}(k_za_z)-1)} & {2jK\xi \; \text{sin}(k_za_z)}\\
   &   \\
{-2jK\xi \; \text{sin}(k_za_z) }&{2G((\text{cos}(k_za_z)-1)}\\
     & {-\xi^2(\text{cos}(k_za_z)+1))}
\end{array}\right)\,,
\end{multline} 
and
\begin{equation}\label{eq11}
B=\left (\begin{array}{cc} {M} & {0}\\
{0}&{J}\end{array}\right)\,.
\end{equation} 
The parameters of these equations are determined numerically: $M=6.2483 \times 10^{-5}$ \si{Kg}, $K= 8.33 \times 10^{3}$ \si{N.m^{-1}}, $G=68 \times 10^{-2}$ \si{N.m.rad^{-1}}, $J=1.0497 \times 10^{-7}$ \si{kg.m^2}, and $\xi=1.3 \times10^{-4}$ \si{m}. We solved the eigenvalue problem theoretically by sweeping $k_z$ in the first irreducible Brillouin zone $(0\leq k_z\leq \pi/a_z)$. A comparison of the results of the theoretical model and the numerical computations for the longitudinal-twist coupling is displayed in \autoref{Figure 3}(i).
\section{Results and discussion}
\begin{figure}[ht]
    \centering
    \includegraphics[width=8.50cm,angle=0]{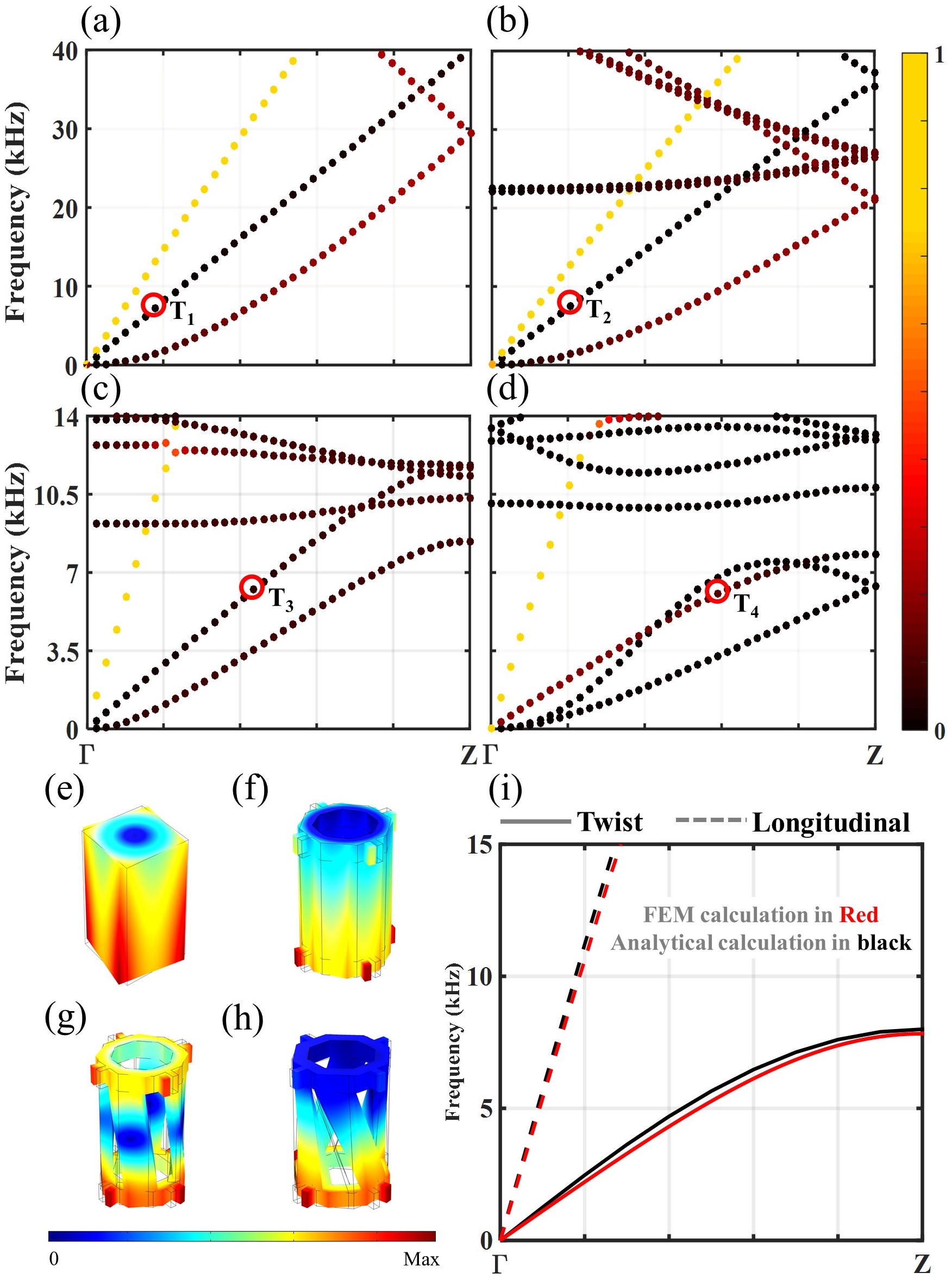}
    \caption{\label{Figure 3}Beam's phononic dispersion diagram in the first irreducible Brillouin zone ($\Gamma Z$) along the $z$-direction for the four structures: (a) the homogeneous unit cell; (b) the octagonal tube unit cell; (c) the non-chiral unit cell; and (d) the chiral unit cell. The color palette indicates the normalized intensity of the polarization along the $z$-direction for each eigenmode. (e), (f), (g), (h) display the modulus of the total displacement for each of the four configurations at the frequencies indicated by the red circles, $T_1$, $T_2$, $T_3$, $T_4$, respectively. (i) The twist (line) and longitudinal (dashed) curves were computed using FEM (red) and theoretical modeling (black).}
\end{figure}
\indent The dispersion diagrams of the four tetragonal-beam cells previously introduced are depicted in \autoref{Figure 3}. The color bar represents the normalized value of the displacement $u_z$ in relation to the total displacement in the unit cell across all band structures. There are two degenerated modes in the case of the homogeneous medium cell (\autoref{Figure 3}(a)): the transverse modes on the one hand, and the twist and longitudinal modes on the other. According to the polarization of $u_z$, the propagation of the longitudinal wave used for excitation has no effect on the mode of twist, as the corresponding mode of the latter remains inactive (black colored). In contrast, the longitudinal mode has the highest intensity (yellow colored). Similarly, for the octagonal shaped cell, the twist mode stays inactive while the longitudinal wave propagates, as illustrated in \autoref{Figure 3}(b). In addition, the two transverse modes in the non-chiral cell remain degenerated (\autoref{Figure 3}(c)) and T-L conversion does not occur in this structure neither. Finally, the chiral cell undergoes a degeneration lift for the transverse modes: a phenomenon commonly referred to as acoustic activity, which is typically triggered by the cell's chirality. Besides, \autoref{Figure 3}(d) shows that a portion of the energy propagating longitudinally has been converted into twist, as indicated by the polarization color of the twist mode turning red. The T-L conversion analysis of the four-cells is further enhanced by depicting screenshots of the modulus of the displacement vector field at a specific frequency, as overlaid in \autoref{Figure 3}(e), (f), (g), and (h). In summary, for the first three unit cells: the homogeneous, octagonal tube, and non-chiral cells, the polarisation of the twist mode is inactive along the $z$-direction. Nevertheless, the mode of twist associated with the chiral unit cell has increased in intensity, indicating that the longitudinal energy has been converted into twist motion around $z$-axis. Thus, chirality within the structure of the metamaterial is a prerequisite for the L-T conversion.
\begin{figure}
    \centering
    \includegraphics[width=8.50cm,angle=0]{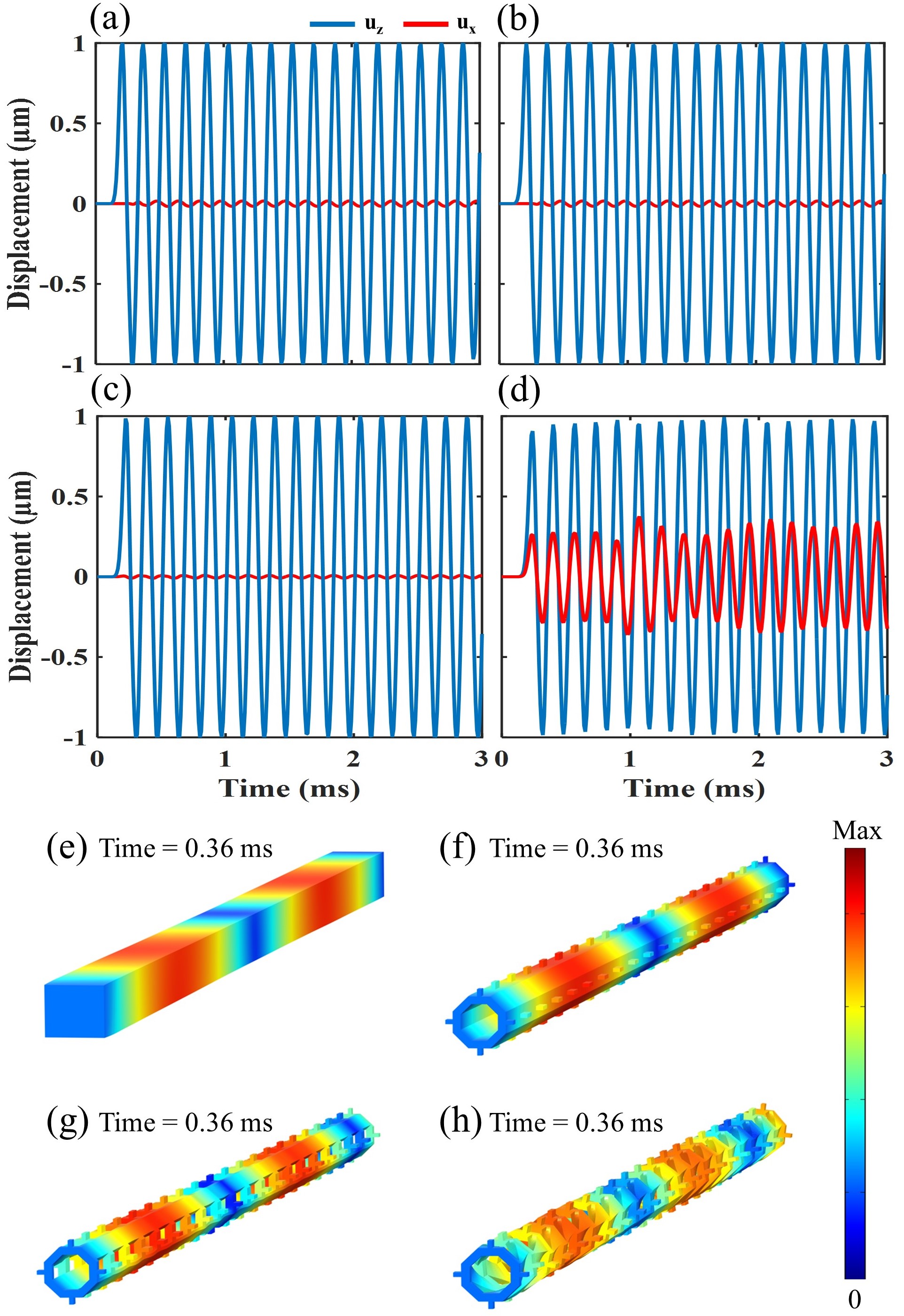}
    \caption{\label{Figure 4}Time-dependent examination of the $u_z$ and $u_x$ displacements for each of the configurations stacked in an array of $20$ cells: (a) A homogeneous cell. (b) A unit cell shaped like an octagon. (c) A non-chiral cell. (d) A chiral cell. (e), (f), (g), and (h) Screenshots of the total displacement in the array created by the previous configurations, respectively, at time 0.36 \si{ms}.}
\end{figure}\\
\indent\autoref{Figure 3}(i) displays the numerical and theoretical dispersion curves evaluated for the longitudinal and twist modes in the first Brillouin zone without accounting for the transverse modes to demonstrate that the theoretical model appropriately captures the chiral cell's longitudinal-twisting coupling. The findings of the theoretical and numerical models are overlaid in black and red, respectively. For both models, the dashed curves indicate the longitudinal modes while the lined curves correspond to the twist modes. This study reveals that the numerical and theoretical dispersion curves are in excellent agreement.\newline
\indent \autoref{Figure 4} highlights the dynamic aspect of the L-T conversion, which we explored through a time-varying analysis of the displacements $u_x$ and $u_z$ at the output face of the array. The $u_z$ displacement recorded at output in all four configurations follows the same trend as the excitation source except for a slight phase shift acquired during the propagation of the wave from the input to output. In regards to $u_x$ displacement, the three structures (\autoref{Figure 4}(a), (b), and (c)) have experienced the Poisson effect, resulting in very small $u_x$ displacements. Notwithstanding, the chiral cell exhibited a static L-T conversion effect, as indicated by the increase in $u_x$ at the expense of $u_z$ (\autoref{Figure 4}(d)), implying that the excitation wave has been converted from longitudinal to an L-T combination.\newline
\indent This results in a characteristic conversion of a $6$ \si{kHz} longitudinal wave propagating through an array of $20$ chiral cells into a wave that contains both longitudinal and twist modes, which can be explained by the dynamic coupling between compression and twist at this frequency. The slight trend difference between the $u_x$ and $u_z$ displacements is due to the fact that the speed of the longitudinal wave is greater than the twist wave, as evidenced by the difference in their slopes in the phononic dispersion diagram (\autoref{Figure 3}(d)). To clarify the wave conversion process, screenshots were captured after $0.36$ \si{ms} of the excitation of wave propagation in the four structures, as shown in \autoref{Figure 4}(e), (f), (g), and (h).\newline
\begin{figure}
    \centering
    \includegraphics[width=8.50cm,angle=0]{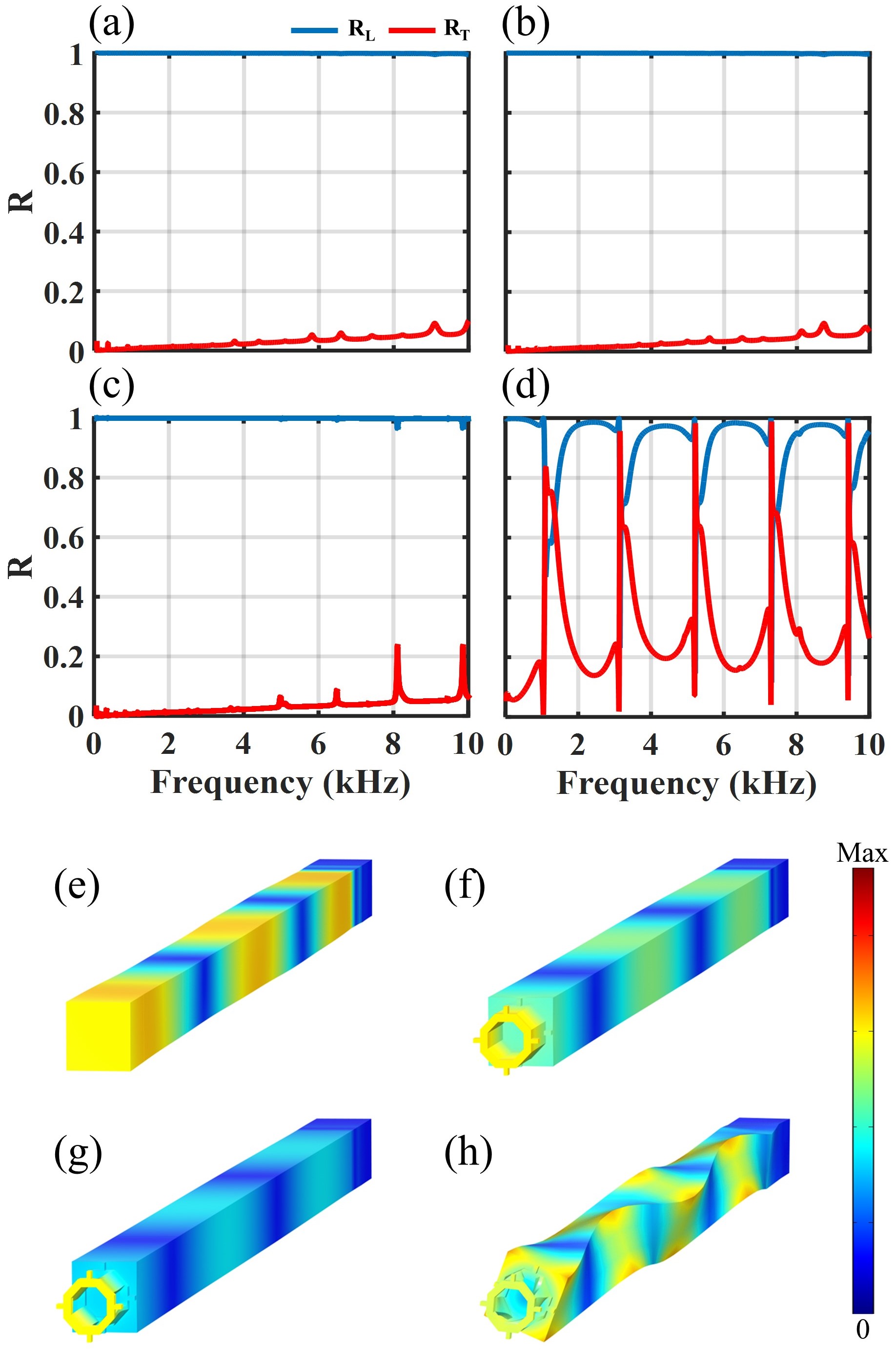}
    \caption{\label{Figure 5}Longitudinal and twist ratios as a function of frequency for the four structures. (a) The homogeneous unit cell. (b) The unit cell shaped like an octagonal tube. (c) The non-chiral unit cell. (d) The chiral unit cell. Screenshots of the total displacement in the second medium and the unit cells at a frequency of 7434.9 Hz. (e) The homogeneous unit cell. (f) The unit cell shaped like an octagonal tube. (g) The non-chiral unit cell. (h) The chiral unit cell.}
\end{figure}
\indent \autoref{Figure 5} elucidates the conversion relationship between the longitudinal and transverse conversion ratios ($R_L$ and $R_T$) as a function of frequency. The longitudinal ratio, conspicuously, outperforms the twist ratio by many orders of magnitude for the first three cells (\autoref{Figure 5}(a), (b), and (c)) across the entire frequency range of $20$ \si{Hz} to $10$ \si{kHz}, thereby indicating that the wave retains its longitudinal polarization and does not undergo any twist conversion. In contrast, \autoref{Figure 5}(d) illustrates that for the chiral unit cell there is a significant increase in $R_T$ within the overall range of frequencies depicted. The $R_L$ decreased to a minimum value at $1.1$ \si{kHz} while the $R_T$ increased. Besides, in the frequency ranges where $R_T$ is maximal, there is a quasi-total conversion of the longitudinal wave into twist, or at least it is partially converted for the other frequencies where the waveform will consist of a combination of longitudinal and twist polarizations. This confirms the alteration of at least a part of the longitudinally polarized wave and, therefore, the credibility of the occurrence of the L-T conversion effect within the spanned frequency range.\newline
\indent The screenshots overlaid in \autoref{Figure 5}(e), (f), (g), and (h) highlight the total displacement field distributions at the frequency of $7434.9$ \si{Hz} in the medium $2$, within which we evaluated the conversion ratios. Moreover, \autoref{Figure 5}(h) clearly shows the occurrence of the L-T conversion in the chiral cell, unlike the first three images.

\section{Conclusion}
\indent The potential of twist metamaterials in the L-T conversion has been demonstrated by comparing multiple configurations, namely: a homogeneous medium, an octagonal-tube, a non-chiral, and a chiral cell. We have demonstrated the effect of chirality within the structure of twist metamaterials on the dynamic conversion of longitudinal into twist waves. The numerical model based on FEM and the theoretical model based on Lagrange's equations for a spring-mass system are in good agreement. Moreover, we have shown that twist metamaterials can operate over a wide range of frequencies to convert longitudinal waves into a combination of longitudinal and twist or even quasi-pure twist waves at specific frequencies, as evidenced by the time-dependent analysis as well as the harmonic study of the L-T conversion ratios. As an engineering application, the proposed approach may be used in the design of numerous new devices, including ultrasonic transducers, ultrasonic foraging, or ultrasonic welding.\newline

\section*{Acknowledgements}
M.K. is grateful for the support of the EIPHI Graduate School (Contract No. ANR-17-EURE-0002) and by the French Investissements d’Avenir program, project ISITEBFC (Contract No. ANR-15-IDEX-03).

\bibliography{Mybib}

\end{document}